# Transfer Deep Reinforcement Learning-enabled Energy Management Strategy for Hybrid Tracked Vehicle

Xiaowei Guo[1*], Teng Liu[2*], Bangbei Tang[3], Xiaolin Tang[2], Jinwei Zhang[4], Wenhao Tan[2], and Shufeng Jin[2]

[1]School of Robot Engineering, Yangtze Normal University, Fuling, 408100, China
[2]College of Automotive Engineering, Chongqing University, Chongqing, 400044, China
[3]School of intelligent manufacturing engineering, Chongqing University of Arts and Sciences, Chongqing, 402160, China
[4]Department of Mechanical and Mechatronics Engineering, University of Waterloo, N2L 3G1, Canada

* Equal contribution. Corresponding author: Bangbei Tang (e-mail: tangbangbei@126.com), Xiaolin Tang (e-mail: tangxl0923@cqu.edu.cn)

This work was supported by Chongqing Science and Technology Project under grant number cstc2019jcyj-msxmX0636 and cstc2019jcyj-msxm2508, Research Startup Project of Yangtze Normal University under grant number 2016KYQD16, and Chongqing Education Commission Project under grant number KJ1712297 and KJQN20191321.

**ABSTRACT** This paper proposes an adaptive energy management strategy for hybrid electric vehicles by combining deep reinforcement learning (DRL) and transfer learning (TL). This work aims to address the defect of DRL in tedious training time. First, an optimization control modeling of a hybrid tracked vehicle is built, wherein the elaborate powertrain components are introduced. Then, a bi-level control framework is constructed to derive the energy management strategies (EMSs). The upper-level is applying the particular deep deterministic policy gradient (DDPG) algorithms for EMS training at different speed intervals. The lower-level is employing the TL method to transform the pre-trained neural networks for a novel driving cycle. Finally, a series of experiments are executed to prove the effectiveness of the presented control framework. The optimality and adaptability of the formulated EMS are illuminated. The founded DRL and TL-enabled control policy is capable of enhancing energy efficiency and improving system performance.

## I. INTRODUCTION

The rise of fuel vehicle production and ownership has aroused widespread concerns about the shortage of energy sources and environmental pollution. Electric vehicles (EVs) are regarded as a promising solution to address the consumption of fossil fuel and emissions. However, limited power battery capacity and the inconvenience of charging operation block the further promotion of EVs [1], [2]. Hybrid electric vehicles (HEVs) make a tradeoff between the travel range and energy economy as well as emissions, which can also realize fast energy supplement through fuel filling. Given that, HEVs have been actively investigated around the world [3], [4], and have become an attractive alternative to fuel vehicles and EVs under the current technology level.

HEVs involve two or more energy sources, so there is a need for energy management strategies (EMSs) to distribute power supplement among several power sources for energy efficiency improvement and emissions reduction [5]. Generally speaking, EMSs can be divided into three types: rule-based methods, optimization-based techniques, and learning-based approaches.

Rule-based strategies rely on predefined rules instead of prior knowledge of driving situations [6], and it is widely welcomed among the current HEVs because of low computational cost and memory space [7]. However, rule-based strategies, which are not adaptive to varied driving conditions and vehicle types, require abundant human experiences and high calibration efforts [8]. Optimization-based strategies are considered as practical alternatives which can obtain the optimal control through the known or predicted driving conditions [9]. Some methods can achieve global optimization, such as dynamic programming (DP) [10], Pontryagin minimum principle (PMP) [11], genetic algorithm (GA) [12], game theory (GT) [13] and convex programming (CP) [14], and some methods can provide instantaneous optimization, such as equivalent consumption minimization strategy (ECMS) [15], model predictive control (MPC) [16] and stochastic dynamic programming (SDP) [17]. However, complicated traffic situations and uncertain driver preferences bring difficulties to the real-time application of optimization-based strategies.

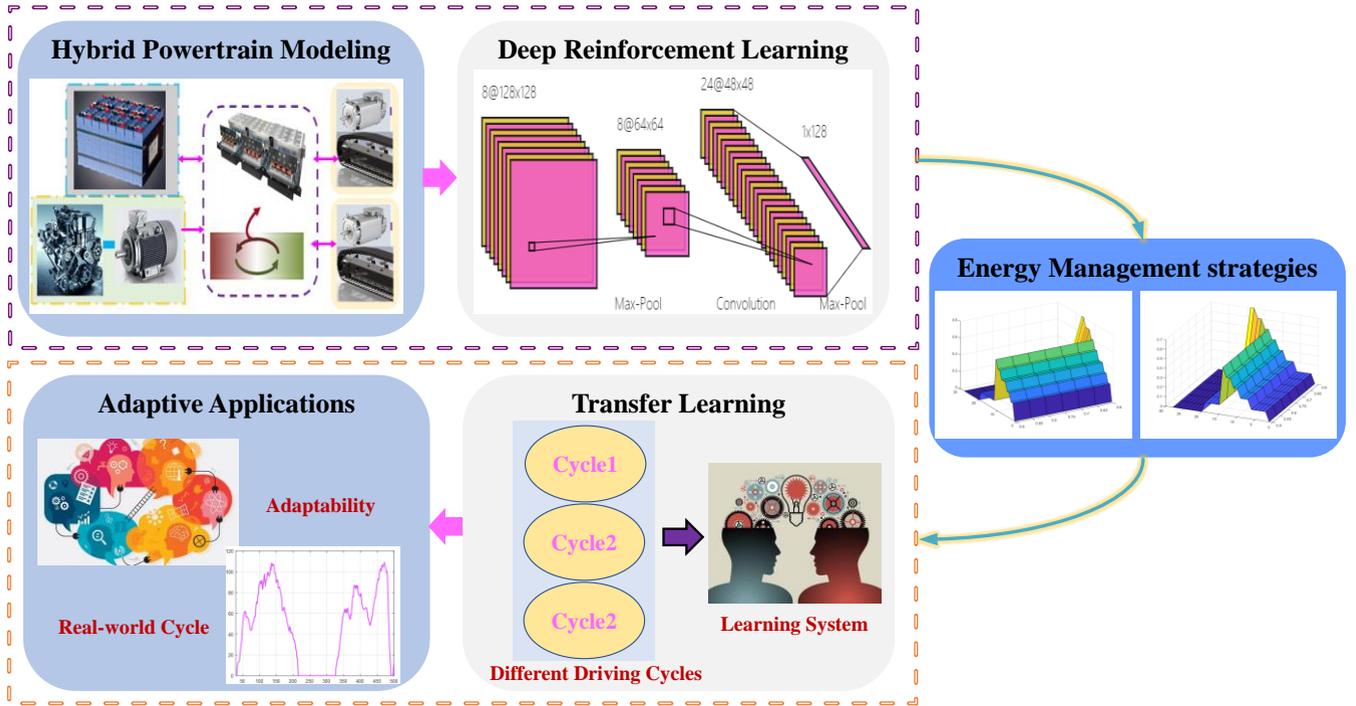

**FIGURE 1.** Adaptive energy management control framework based on DRL and TL for hybrid powertrain.

With the rapid development of artificial intelligence and computing power, learning-based strategies are gradually applied in the energy management field of HEVs. Learning-based strategies get rid of the dependence of accurate prior travel information and make use of historical driving knowledge to learn optimal control instead. For example, Liu T. et al. [18] proposed a reinforcement learning (RL)-based adaptive energy management (RLAEM) for a hybrid electric tracked vehicle (HETV), and RLAEM showed better performance in adaptability, optimality, and learning ability than SDP in simulations. Qi X. et al. [19] applied an RL-based real-time EMS for plug-in HEVs (PHEVs), which realized a balance between real-time performance and optimal energy savings. In [20], better power distributions between the battery and the ultracapacitor of PHEVs were obtained through the RL-based method, and 16.8% energy loss reduction was achieved. However, the discrete state space and action space of RL hinder its further application in EMS of HEVs, and the emergence of deep reinforcement learning (DRL) has bridged over this difficulty.

The successful application of DRL in Alpha GO [21] triggered the enthusiasm of researchers from various fields, including the energy management of HEV. Wu J. et al. [22] employed deep Q-learning (DQL) for energy management issues of a hybrid electric bus (HEB), and the results indicated that DQL-based strategy outperforms Q-learning in training time and convergence rate. Du G. et al. [23] used a new optimization method (AMSGrad) to upgrade the weight of the neural network in DQL and improved training speed and fuel economy. Han X. et al. [24] put forward a double deep Q-learning-based control structure, which can prevent the training process from falling into the over-optimistic estimate of policy value compared with conventional DQL. Since the DQL cannot output continuous actions, Tan H. and Wu Y. et al. [25], [26] proposed energy management strategies based on DDPG and simulation showed that DDPG-based strategy performs better than Q-learning and achieves results closed to DP. However, the real-time application of DRL in energy management hit a bottleneck at the moment, for the current training efficiency of DRL remains at a low level.

This paper aims to address the defect of DRL in tedious training time. By combining DRL and the transfer learning (TL), an adaptive energy management strategy for HETV is proposed, as depicted in Fig. 1. First, the powertrain model of a hybrid electric tracked vehicle (HETV) is constructed, and the energy management problem is formulated. Then, a bi-level control framework is established to derive EMSs, which combines DDPG and TL. Finally, the effectiveness of the presented control framework is verified through a series of experiments.

This paper has three main contributions: (1) an adaptive energy management policy based on DDPG and TL is proposed for a HETV; (2) a bi-level control framework is founded to address the optimization control problem, which includes the driving cycle classification and DRL; (3) the optimality and adaptability of the presented method are evaluated in detail via a series of simulation experiments.

The rest of this paper is arranged as follows. In Section II, the powertrain modeling of HETV and the energy management problem are presented. Section III describes the upper-level application of DDPG to train EMS in different speed intervals and low-level employment of TL to transform the pre-trained neural networks for a novel driving cycle. In Section IV, a series of experiments is established to demonstrate the effectiveness of the proposed bi-level energy management strategy with its results analyzed. Finally, conclusions are described in Section V.

## II. DYNAMICAL MODELING AND CONTROL PROBLEM

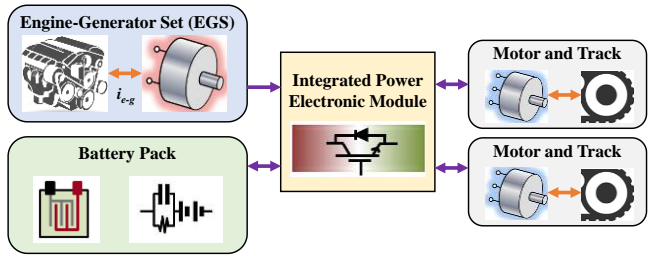

FIGURE 2. Configuration of the studied HETV.

The powertrain configuration of this research is a series HETV, as shown in Fig. 2. The vehicle powertrain mainly consists of a battery pack, an engine-generator set (EGS), and two traction motors. In this section, the vehicle demand power is derived firstly. Then, the detailed modeling of the battery and EGS are illustrated, respectively. Finally, the energy management problem is formulated. The key parameters of HETV are listed in Table I.

TABLE I
DEFAULT POWERTRAIN PARAMETERS OF THE HETV.

| Keyword | Value | Unit |
|---|---|---|
| Curb weight $m$ | 2500 | kg |
| Generator Inertia $J_g$ | 0.1 | kg·m$^2$ |
| Engine Inertia $J_{eg}$ | 0.2 | kg·m$^2$ |
| Fixed transmission ratio $i_{e-g}$ | 1.2 | / |
| Electromotive force parameter $K_e$ | 0.8092 | Vsrad$^{-2}$ |
| Electromotive force parameter $K_x$ | 0.0005295 | NmA$^{-2}$ |
| Width of contacting track $l$ | 1.53 | m |
| Air resistance coefficient $C_d$ | 0.9 | / |
| Windward area $A$ | 3.4 | m$^2$ |
| Vehicle tread $B$ | 1.42 | m |
| Rolling resistance coefficient $\mu$ | 00494 | / |
| Minimum State of Charge $SOC_{min}$ | 0.5 | / |
| Maximum Sate of Charge $SOC_{max}$ | 0.9 | / |
| Battery rated capacity $C_b$ | 37.5 | Ah |

### A. POWER DEMAND

The power demand of HETV is notably different from the wheeled vehicle. In the wheeled vehicle, the tractive power is heading power. But in a tracked vehicle, the propelled power of HETV includes heading power and steering power:

$$\begin{cases} P_r = (F_i + F_j + F_\mu)v + M\omega \\ v = \dfrac{v_1 + v_2}{2} \\ \omega = \dfrac{v_1 - v_2}{B} \end{cases} \quad (1)$$

where $F_i$ is the air drag, $F_j$ is the inertial force, $F_\mu$ is rolling resistance, and the slope resistance is ignored in this research. $M$ indicates the force moment of resistance, $\omega$ is the steering angular velocity. The $v$, $v_1$, $v_2$ are the average velocity and speed of the two tracks, respectively, $B$ is the vehicle tread. The aforementioned resistance forces can be obtained by the following equations:

$$\begin{cases} F_j = ma \\ F_i = \dfrac{C_d A}{21.15} v^2 \\ F_\mu = \mu mg \end{cases} \quad (2)$$

where $C_d$ is the coefficient of air resistance, $A$ is the windward area of the vehicle, $m$ indicates the vehicle mass, $a$ is the vehicle acceleration, $g$ is the gravitational acceleration, and $\mu$ is the coefficient of rolling resistance. The $M$ can be calculated by the empirical equation as below [27]:

$$\begin{cases} \lambda_t = \lambda_{max}(0.925 + 0.15\dfrac{R}{B})^{-1} \\ M = 0.25\lambda_t mgl \\ R = \dfrac{v}{|\omega|} \end{cases} \quad (3)$$

where $R$ is the turning radius, $\lambda_t$ indicates the lateral resistance coefficient, $\lambda_{max}$ is the maximal value of the lateral drag coefficient. $l$ is the width of the contacting track.

During the process of vehicle running, the tractive power is provided by EGS and battery. Considering the loss efficiency of energy transition, the power demand at the wheels is acquired by:

$$P_r = (P_g \eta_g + P_b)\eta_m^i \quad (4)$$

where $P_g$ is the power of a generator, $P_b$ indicates the power of battery, $\eta_g$ and $\eta_m$ represents the efficiency of generator and electric motor, respectively. When the motor work to propel the vehicle, $i = 1$, otherwise, $i = -1$, work as a generator to recover the braking energy.

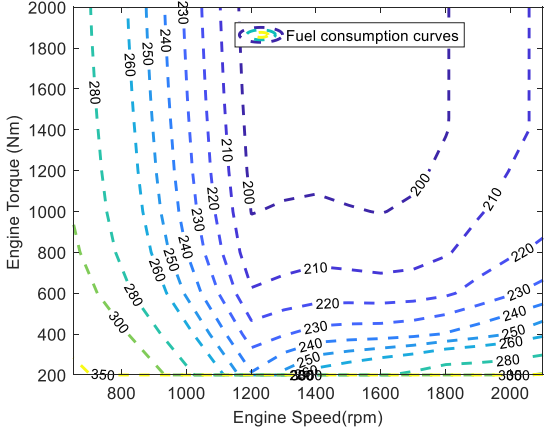

**FIGURE 3.** Brake-specific fuel consumption curves of engine in the research HETV.

### B. POWERTRAIN MODELING

According to the human experience, the engine, motor, and generator are modeled by their corresponding efficiency map, respectively. As show in Fig 3, the equivalent fuel consumption rate of engine can be acquired by, $fuel(t)$ = interpolation ($T_{eg}$, $n_{eg}$), $fuel(t)$ is the equivalent fuel consumption, interpolation indicates the interpolation function, $T_{eg}$, $n_{eg}$ represent the torque and rotational speed, respectively. Similarly, the efficiency of a motor can be obtained obtain by $\eta_m$ =interpolation ($T_m$, $n_m$), $\eta_m$ is the efficiency of the motor, $T_m$, and $n_m$ indicate the torque and rotational of the motor, respectively.

Generator model: Based on the empirical equivalent electric circuit of EGS, the dynamic parameters of the generator can be derived from follows equations:

$$\begin{cases} U_g = K_e n_g - K_x n_g I_g \\ T_g = K_e I_g - K_x I_g^2 \\ \frac{2\pi}{60}(\frac{J_{eg}}{i_{e-g}^2} + J_g)\frac{dn_g}{dt} = \frac{T_{eg}}{i_{e-g}} - T_g \\ n_{eg} = n_g / i_{e-g} \\ \pi K_x = 3PL^g \\ P_g = U_g I_g \end{cases} \quad (5)$$

where $U_g$, $I_g$, and $P_g$ indicate the voltage, current, and power of the generator, respectively. $T_{eg}$, $T_g$, $n_{eg}$, and $n_g$ are the torque and rotational speed of engine and generator, respectively. $K_e$ is the electromotive force coefficient, $K_x n_g$ indicates the electromotive force. $J_{eg}$ and $J_g$ represent the inertia moment of engine and generator, respectively. $i_{e-g}$ represents the fixed transmission ratio between engine and generator. $P$ is the count of poles and $L^g$ is the synchronous inductance of the armature.

Battery model: the battery is very important in the powertrain. It can provide power to drive the system when the vehicle runs in high power demand and recycle the braking energy when the vehicle is decelerated. There are many dynamic parameters of the battery, such as terminal voltage, state of charge (SOC), capacity, internal resistance and so on. One of the most important factors is SOC, it indicates the current amount of electric charge stored in the battery, and it also is one of the state variables in this research. In this paper, the battery modeling adopts the internal resistance model, and SOC can be calculated as follows:

$$\begin{cases} U_b = V_{oc} - I_b r_b \\ \dot{SOC}(t) = \frac{(V_{oc} - \sqrt{V_{oc}^2 - 4r_b P_b})}{2C_b r_b} \\ P_b = U_b I_b \end{cases} \quad (6)$$

where $P_b$, $U_b$, $V_{oc}$, $I_b$, and $r_b$ are the power, voltage, open-circuit voltage, the current, and internal resistance of the battery. $C_b$ is the rated capacity, and the influence of temperature and battery aging is ignored.

### C. ENERGY MANAGEMENT PROBLEM

The object of EMS is to reduce fuel consumption and sustain the SOC at the appropriate interval. To realize this, the reasonable action variable, state variable, and cost function should be designed. The continuous engine power is selected as the sole action variable in this paper, the state variable consists of vehicle velocity, vehicle acceleration, and battery SOC. To address the energy management problem, the reward function is defined as follows:

$$\begin{cases} reward = -[\alpha \bullet fuel(t) + \beta(\Delta SOC)^2] \\ \Delta SOC = \begin{cases} SOC(t) - SOC_r & SOC(t) < SOC_r \\ 0 & SOC(t) \geq SOC_r \end{cases} \end{cases} \quad (7)$$

where $fuel(t)$ represents the instantaneous fuel consumption of the engine. $\alpha$ and $\beta$ are two positive weight factors, their values should be tuned appropriately to improve the performance of fuel consumption as much as possible. $SOC_r$ is a reference value of battery SOC to achieve charging sustaining. To guarantee the components operate safely and reliably, the following inequalities should be observed:

$$\begin{cases} T_{eg,\min} \leq T_{eg}(t) \leq T_{eg,\max} \\ n_{eg,\min} \leq n_{eg}(t) \leq n_{eg,\max} \\ n_{g,\min} \leq n_g(t) \leq n_{g,\max} \\ P_{r,\min} \leq P_r(t) \leq P_{r,\max} \\ n_{m,\min} \leq n_m(t) \leq n_{m,\max} \\ SOC_{\min} \leq SOC(t) \leq SOC_{\max} \end{cases} \quad (8)$$

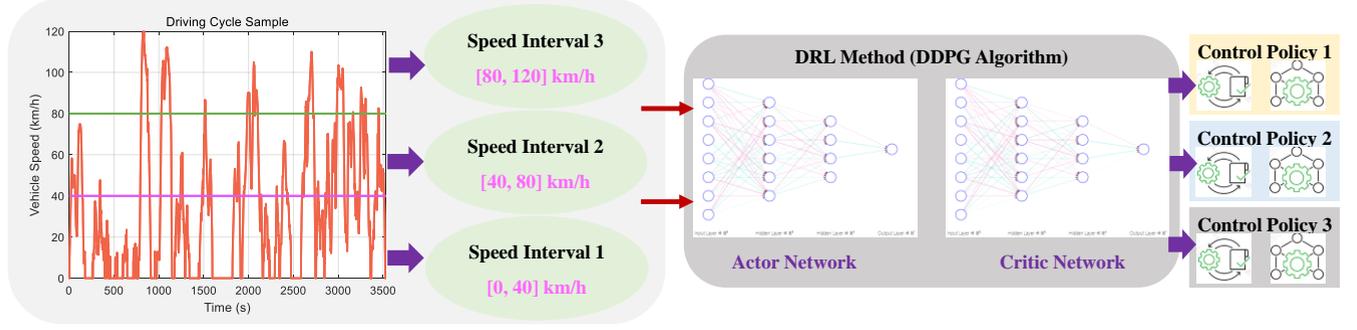

**FIGURE 4.** Speed classification for energy management policy generation using DRL approach.

where $T_{eg}$ and $n_{eg}$ are the torque and rotation speed of the engine, $n_g$ and $n_m$ represent the rotational speed of generator and motor, respectively. $P_r$ is the power demand, min and max indicate the minimum value threshold and maximal value threshold for the corresponding variable.

## III. TRANSFER DEEP REINFORCEMENT LEARNING

In the next part, a bi-level control structure is designed to address the energy management problem for an HETV, and the training process is introduced, firstly. The driving cycles are classified into three-speed intervals to learn the EMS. Then, the DQN and DDPG are introduced to search for the optimal control policy for the studied powertrain. Finally, transfer learning is employed to speed up the convergence of the training process.

### A. A BI-LEVEL FRAMEWORK

As shown in Fig. 1, the upper-level is a DDPG-based control framework, which is utilized to derive an optimal control strategy. The vehicle speed dataset is collected from the real driving conditions. To reduce the training time and improve the control accuracy, the vehicle velocity is divided into three-speed intervals that are [0-40], [40-80], [80-120], and they represent low, medium and high speed, respectively, the unit is Km/h. Then, the classified speed intervals are adopted to train the DDPG algorithm separately until the algorithm converges, the trained neural network is stored, as depicted in Fig. 4. The lower level of the control framework is the online application of the obtained EMSs, the TL method is employed to transfer the pre-trained neural network to a new driving cycle. In this way, it takes very few seconds to derive an optimal control strategy for a novel driving cycle.

### B. ALGORITHM INTRODUCTION

RL algorithm: RL is a self-learning method by the interaction between agent and environment, it can derive an optimal control strategy for a sequential problem through a trial-and-error method to maximize the accumulated reward. The main part of the RL algorithm includes an agent and an environment. At every time step $t$ ($t = 0,1,2,3\ldots n$), the agent observes a state $s_t$ ($s_t \in S$, $S$ includes all state variables) from the environment, then action $a_t$ is chosen from $A$ ($A$ is a set of all action variables) based on policy $\Pi$ and implemented to the environment, then the environment transfer into a next state $s_{t+1}$ and an instant reward $r_{t+1}$ feedback to the agent, and so on until the last sate. The optimal control policy is a map from current state to an optimal action, which can be obtained by maximizing the total discounted expected reward as follows [28]:

$$R_t = \sum_{t=0}^{\infty} \gamma^t r_{t+1} \qquad (9)$$

where $\gamma$ indicates the discount rate, $\gamma \in [0,1]$. The discount rate adjusts the importance between immediate and future reward, and guarantee the convergence of the optimal policy. Generally, with the state and action space increase, the conventional RL algorithm is easily caught in the problem of "curse of dimensionality" and the calculation time increases exponentially.

DRL algorithm: the DRL algorithm is proposed to address the above problems. DRL is the combination of conventional RL and neural networks. The conventional RL algorithm acquires the optimal control policy through update the Q-value table, it requires a lot of memory to store the corresponding Q-value when the spaces of action and state are numerous. The DRL algorithm is utilized the neural network to estimate the Q-value, and the already trained neural network can be employed as a black box. The famous DRL algorithm is DQN, which consists of two neural networks called target-network and evaluate-network, respectively. State variables are the inputs of evaluate-network, and the outputs are the Q-value of corresponding action variables, the input of target-network is the next state variable, and the output is the target Q-value. The update method of DQN as follows:

$$\begin{cases} Q(s,a;\theta) = Q(s,a;\theta) + \delta(y_t - Q(s,a;\theta)) \\ y_t = r + \gamma \max_{a'} Q(s',a';\theta') \end{cases} \qquad (10)$$

where $\delta$ indicates the learning rate, $\gamma$ is the discount rate, $\theta$ are the parameters of the evaluate network, $\theta'$ are the target network parameters. $y_t$ represents the target Q-value, loss function of the network as follows:

TABLE II
THE PSEUCUDOCODE OF DQN

| Algorithm: DQN |
|---|
| Randomly initialize θ, θ′=θ |
| Initialize the reply experience pool capacity with $D$ |
| For every episode do |
|     Reset initial state |
|     For $t$=1, 2, … n do |
|         Choose $a_t$ randomly with probability ε |
|         Choose $a_t$=argmax (Q ($s, a$; θ)) with probability 1-ε |
|         Execute $a_t$ at state $s_t$, observe the $s_{t+1}$ and $r_{t+1}$ |
|         Store ($s_t, a_t, r_{t+1}, s_{t+1}$) in D |
|         Randomly select minibatch of ($s_t, a_t, r_{t+1}, s_{t+1}$) from $D$ |
|         Calculate the target Q-value $y_t$ |
|         Performance a gradient decent on *loss* about θ |
|         Every $k$ steps θ′=θ |
|     End for |
| End for |

$$loss = (r + \gamma \max_{a'} Q(s',a';\theta') - Q(s,a;\theta))^2 \quad (11)$$

the gradient descent and backpropagation are applied to update the parameters of the evaluated network, and the parameters of the target network are replaced by the evaluated network every $k$ steps directly, which called hard update method. Because of the outputs of the DQN are the discrete Q-values, it cannot address the problem with continuous action space. The pseudocode about the training process of DQN as Table II.

DDPG algorithm: DDPG is an actor-critic network, actor, and critic contain an evaluate-network and a target-network, respectively. The inputs of the evaluate network are states and the output is the corresponding action. This method adopts the deterministic policy gradient that means the output of the actor-network is the deterministic action rather than the probability of corresponding actions. Therefore, the DDPG algorithm can address the issue with continuous actions and states. States and corresponding actions are the inputs of the critic network, the output is the corresponding Q-value. To reduce the correlation between data, the replay experience pool method is employed in the training process, and the priority experience replay is applied to faster the training time, the noise technique is taken to increase exploration. Different from DQN, the parameters updated of DDPG is the way of soft update, the loss function and policy gradient as follows:

$$\begin{cases} loss = \frac{1}{n}\sum_t [y_t - Q(s_t,a_t|\theta_c)|_{a_t=\mu(s_t|\theta_a)}]^2 \\ a_t = \mu(s_t|\theta_a) \\ y_t = r + \gamma Q'(s_{t+1},\mu'(s_{t+1}|\theta_a')|\theta_c') \\ \nabla_{\theta_a} loss = \frac{1}{n}\sum_t [\nabla_{a_t} Q(s_t,a_t|\theta_c)\nabla_{\theta_a}\mu(s_t|\theta_a)] \end{cases} \quad (12)$$

where $n$ is the number of minibatch, $\theta_a$ and $\theta_c$ indicate the parameters of the evaluate network for actor and critic,

TABLE III
THE PROCEDURE OF DDPG ALGORITHM

| DDPG algorithm |
|---|
| Initialize the evaluate network parameters randomly $\theta_a, \theta_c$ |
| Initialize target network parameters $\theta_a'=\theta_a, \theta_c'=\theta_c$ |
| Initialize experience pool with $D$ |
| For every episode do |
|     Initialize $s_1$ |
|     For $t$=1, 2, … n do |
|         Choose $a_t=\mu(s_t|\theta_a)$ |
|         Execute $a_t$ at state $s_t$, observe the $s_{t+1}$ and $r_{t+1}$ |
|         Store ($s_t, a_t, r_{t+1}, s_{t+1}$) in D |
|         Select minibatch ($s_t, a_t, r_{t+1}, s_{t+1}$) with priority experience replay from $D$ |
|         Calculate the target Q-value $y_t$, Equation (12), $y_t = r_{t+1}$ when $s_t$ is terminal |
|         Update $\theta_c$ by minimize *loss*, Equation (12) |
|         Update $\theta_a$ by execute a gradient decent on *loss* about $\theta_a$ |
|         Update target network parameters: |
|         (1-τ) $\theta_c'+\tau\theta_c \rightarrow \theta_c'$ |
|         (1-τ) $\theta_a'+\tau\theta_a \rightarrow \theta_a'$ |
|     End for |
| End for |

respectively, $\theta_a'$ and $\theta_c'$ represent the parameters of the target actor and critic network, respectively. $\mu$ is the function to map an action to the homologous state. $r$ indicates the instant reward, $\gamma$ is the discount rate. The parameter update of the critic network is to minimize the *loss* and the parameter update of actor-network is to implement the gradient descent to loss with $\theta_a$. The pseudocode of the DDPG algorithm is indicated in Table III.

### C. TRANSFER LEARNING

The conventional deep learning algorithm is generally employed to settle the problem of the testing data and training data from the same domain. When the condition is not satisfied, it is very expensive and annoying to rebuilt the model and retrain the collected data. Fortunately, the technique of TL is very beneficial to address this problem. As two research problems are similar, the TL could store a majority of parameters in the neural network and recycle them in the new problem.

TL algorithm: there is a source domain $A_x$ and the correspondent learning task $B_x$, an objective domain $A_y$ and the correspondent learning task $B_y$. Transfer learning with the purpose of improving the learning of the target predictive function $f$ in $A_y$ applying the knowledge in $A_x$ and $B_x$, where $A_x \neq A_y$ or $B_x \neq B_y$ [29]. A domain consists of feature space and a marginal probability distribution. According to whether the feature space is same, there are two kinds of TL algorithms, and they are named homogeneous TL (the feature space is the same) and heterogeneous TL (the feature spaces are different). The heterogeneous TL is usually applied in image recognition and language text classification. However, when the source domain is uncorrelated to the target domain, negative transfer will happen and it will result in bad results. The solution of negative transfer is to divide the big data into several segments and choose the data that can bring good training performance for training.

DDPG is a model-free learning method, its learning goal independent from the accurate model of the learning problem. In this research, the DDPG algorithm is applied to address the problem of EMS for HETV. It may consume a long time to construct the training model for a new driving cycle. Inspired by TL thought, the vehicle speeds of the driving cycle are classified into three intervals firstly. Since the learning task and feature space have similar nature for each speed interval, the TL is useful to transform the trained DDPG parameters from one interval for another. The majority of parameters in the neural network are the same, only the parameters of the output layer should be retrained.

As described in Fig. 1, the DDPG algorithm is utilized for training the EMS for different speed intervals and the relevant learned parameters are stored. Since the driving cycles for the HETV have the same feature space and they are correlated with each other. To improve the computation efficiency for a new driving cycle, the trained parameters are transferred to the new cycle at different speed intervals. By doing this, the EMS could be generated efficiently, and its optimality is able to be guaranteed. In the next section, the effectiveness of the DDPG and TL-enabled EMS is evaluated, and the related simulation results are analyzed.

## IV. SIMULATION RESULTS AND ESTIMATION

This section estimates the optimality, adaptability, and calculative efficiency of the proposed adaptive EMS for HETV. The DP and common DDPG-based optimal control policies are treated as benchmark to evaluate the optimality of the DDPG and TL method. Then, DDPG and TL is compared with common DDPG in detail to assess the convergence rate of the Q-value table. Finally, the DDPG and TL-enabled control strategy are applied to the new driving cycles to verify the adaptability.

### A. OPTIMALITY EVALUATION

To identify the optimality of the proposed EMS based on DDPG and TL (DDPG+TL for short), the DP and common DDPG are taken as baseline methods. Since the DP could generate the global optimal control actions, the differences between the presented approach and DP are used to quantify the degree of optimality. The default parameters in DDPG+TL and common DDPG are completely the same. A standard driving cycle called LA-92 is exploited in this comparison.

Fig. 5 depicts the track of the LA-92 cycle and the related SOC trajectories for the three techniques. Eq. (6) describes the SOC is affected by the batter power $P_b$, and thus SOC is indirectly influenced by the control action (engine power). It can be discerned that the SOC in DDPG+TL is close to that in DP, which reflects that their control sequences are also similar. Furthermore, the SOC variations indicate the DDPG+TL is superior to the common DDPG in the output power of the battery.

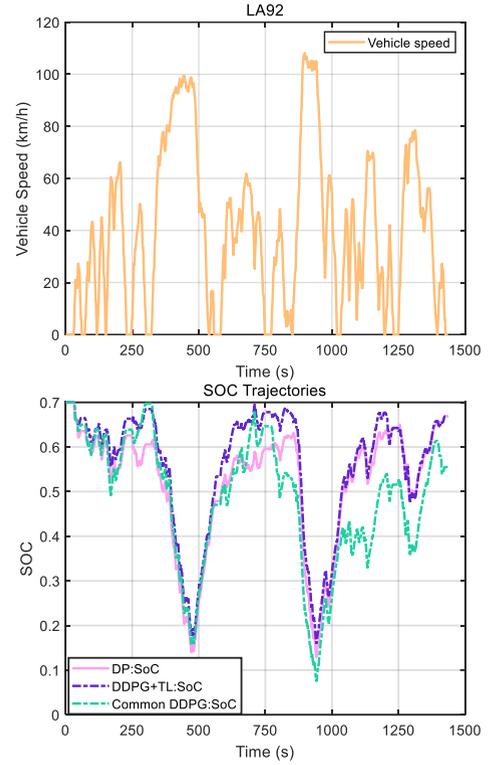

**FIGURE 5.** Standard driving cycle and the relevant SOC curves for three compared methods.

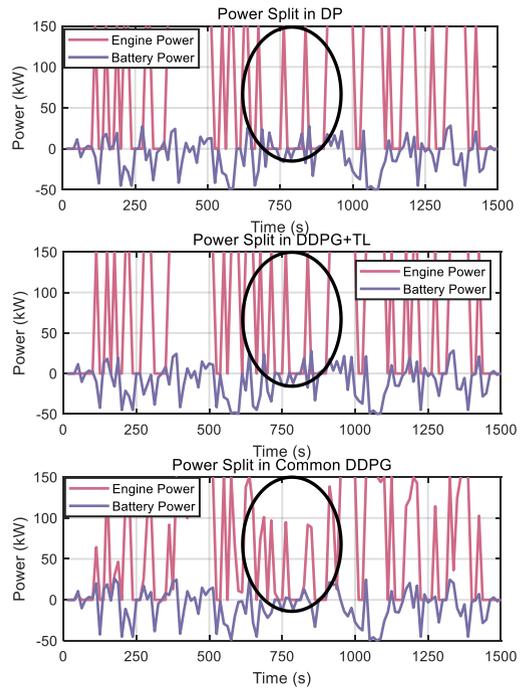

**FIGURE 6.** Power distribution between engine and battery in three control cases.

To show the advantages of the proposed method in control level, the power distributions between engine and battery are analyzed in Fig. 6. The variation trends in DP and DDPG+TL are nearly the same. For concretization, the common DDPG is different from the other two methods in the black circle. This would result in different performance in fuel economy. Hence, in the DDPG+TL method, the transformed learning from the existing energy management knowledge could reduce fuel consumption and guarantee the optimality.

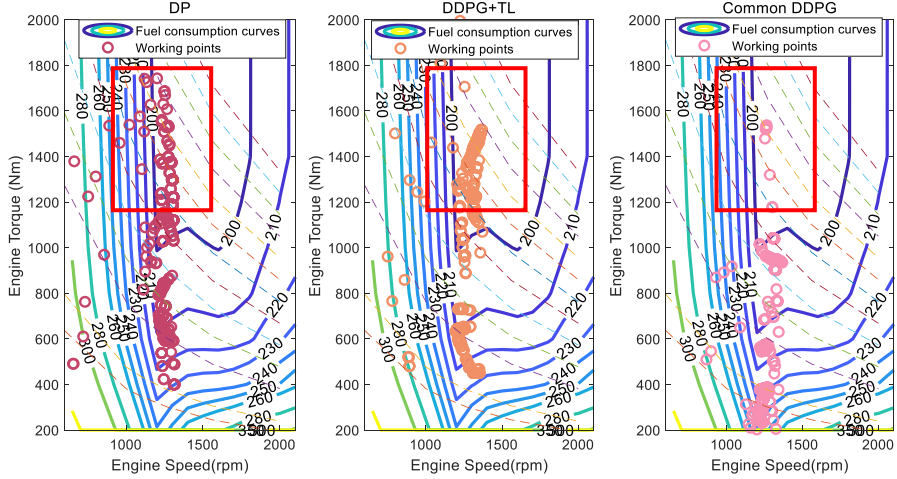

**FIGURE 7.** Working points of the engine in DP, DDPG+TL, and common DDPG for fuel consumption comparison.

The working regions of the engine in these three approaches are displayed in Fig. 7. This comparison is able to exhibit the numeric results in fuel economy. The rectangular box highlights the high-efficiency area of the engine, and the DP and DDPG+TL often operate at this location. It could improve the fuel economy with higher working efficiency in the engine. As a consequence, the quantized fuel consumption is discussed in Table IV. For the same driving cycle, the DDPG+TL achieves a higher fuel economy than the common DDPG and is very close to DP. The above simulation results are capable of demonstrating the optimality of the proposed DDPG+TL method.

TABLE IV
FUEL ECONOMY IN DP, DDPG+TL AND COMMON DDPG.

| Techniques[#] | Final SOC | Fuel Consumption[*] |
|---|---|---|
| DP | 0.6194 | 2.8673 |
| DDPG+TL | 0.6184 | 2.9641 |
| Common DDPG | 0.5051 | 3.2896 |

[#] A 2.30 GHz microprocessor with 31.8 GB RAM was used.
[*] Equivalent fuel consumption, Unit: L/100km.

### B. TRAINING AND LEARNING TIME

The original intention of the DDPG and TL is utilizing the learned neural network parameters to accelerate the searching process of control actions. The sole variable for different EMSs is the driving cycle. Based on the concept of TL, only the output layer of the neural network is retrained for a new driving cycle in the DDPG+TL control case. By doing this, the learning time for a new control policy could be decreased sharply. The DPPG+TL and common DDPG are compared in this part to analyze the convergence rate and training time.

In these two DRL methods, the goal of the neural network is to approximate the Q-value table. Fig. 8 describes the mean error of this Q table along with the training episodes (the number is 1000 in this paper). The downtrends indicate the obtained control sequence becomes better as the episode increases. For each episode, the value of the mean error in DDPG+TL is lower than that in common DDPG. It implies that the proposed technique could learn more knowledge about the environments and thus lead to better control policy. Thus, the DDPG+TL has a faster learning efficiency than the common DDPG.

Another keyword to represent the training efficiency in DRL is the cumulative rewards. As shown in Fig. 9, the accumulated rewards in Eq. (9) increases with the number of episodes. It explains that the energy management controller could realize higher rewards via a trial-and-error procedure. Moreover, the DDPG+TL is also more excellent than the common DDPG with the same episodes. The training time for these two methods on the same driving cycle is depicted in Table V. It is obvious that the DPPG+TL is more efficient than common DDPG because a majority of parameters in the neural network could be recycled. This property enables the proposed EMS to be applied in the real-world driving environment.

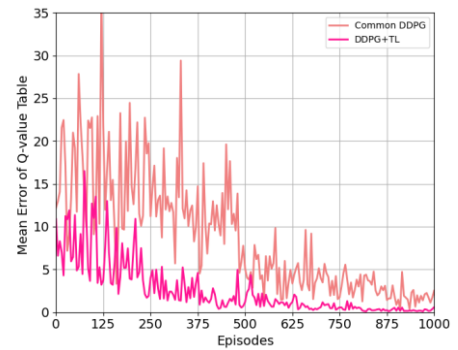

**FIGURE 8.** Mean errors of Q-value table in two DRL techniques.

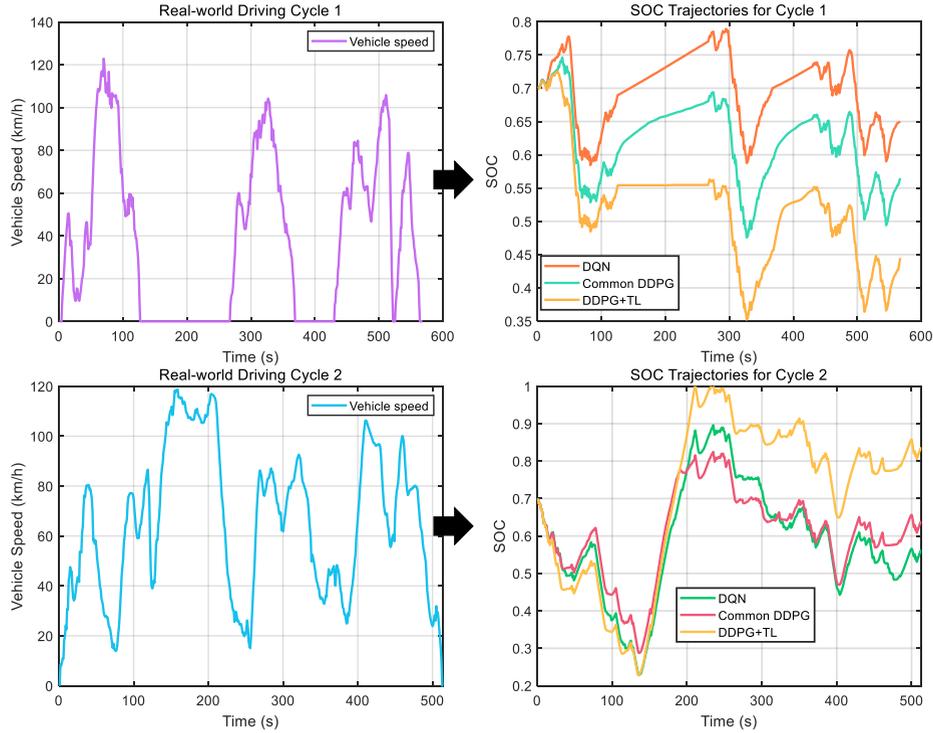

**FIGURE 10.** Two real-world driving cycles and their related SOC trajectories in three DRL methods

**TABLE V**
**TRAINING TIME OF DDPG+TL AND COMMON DDPG.**

| Methods[#] | Training Time (h) |
| --- | --- |
| DDPG+TL | 0.79 |
| Common DDPG | 21.46 |

[#] A 2.30 GHz microprocessor with 31.8 GB RAM was used.

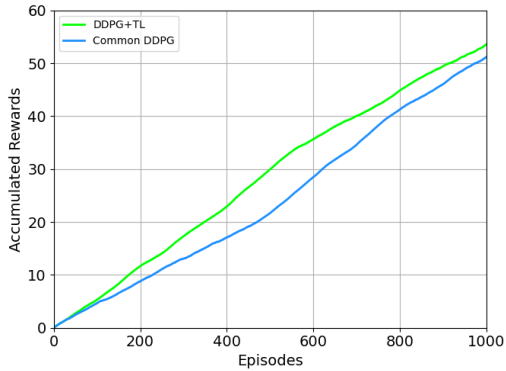

**FIGURE 9.** Cumulative rewards in DDPG+TL and common DDPG for convergence rate comparison.

### C. ADAPTABILITY ESTIMATION

As the presented control framework aims to be adaptive to different cycles by transferring the learned parameters, this subsection validates this structure on two real-world driving cycles. The compared methods are DQN, common DDPG, and DDPG+TL. The trajectories of the vehicle speeds and SOC are displayed in Fig. 10. In this simulation test, these two verified cycles are not included in the learning process. The learned parameters of the neural network are reused on these cycles. The experiment results reveal that the SOC variations are not the same in these three cases. It implies that the output power of the engine (control action) and the working points of the engines are different in these three EMSs.

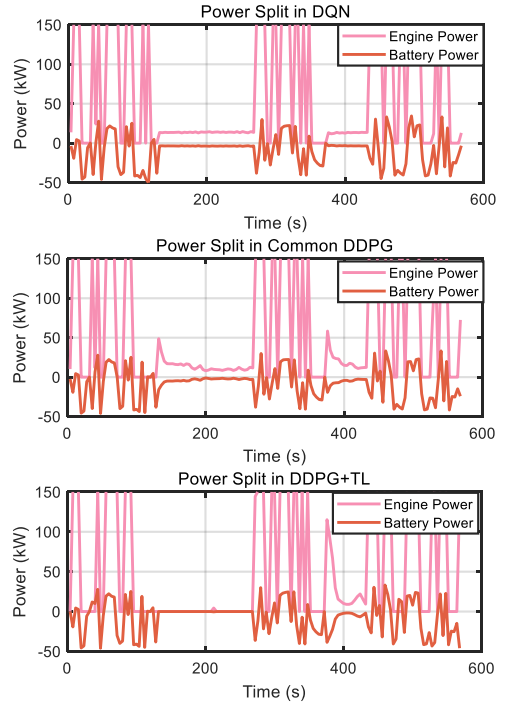

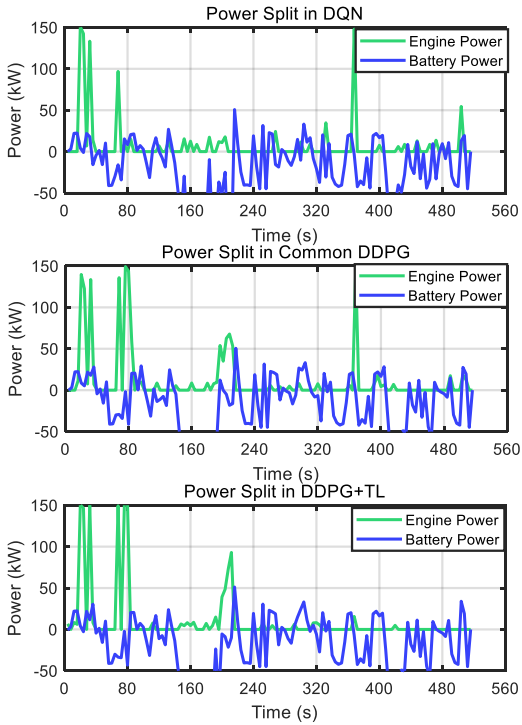

**FIGURE 11.** Power split controls between engine and battery for three methods in two real-world driving cycles: the first figure is cycle 1 and the second figure is cycle 2.

The power-split control curves of engine and battery powers are sketched in Fig. 11. These curves are not the same too, which means the power distributions are different in these three control cases. The different variations of these tracks can be ascribed to the recycled learning experiences, which promote the performance of DDPG+TL. Since there are actor and critic networks in DDPG, the DQN (DQN only has one network) is faster than common DDPG. However, with the help of the TL method, the presented DDPG+TL could achieve the highest learning efficiency and thus result in the higher control effects. Moreover, the proposed method is able to adapt to different driving cycles in energy management problems. It implies this control structure has the potential to be applied in online applications.

## V. CONCLUSION

To address the tedious training time in DRL methods, the work combines the DDPG and TL to derive an adaptive energy management controller for HETV. This control framework is easy to generalize into another hybrid powertrain. The effectiveness of the relevant EMS is evaluated, including the optimality, convergence rate and adaptability. The advantage in training and learning time is exhibited in detail. The proposed method is proven to have the potential to be applied in real-world environments. Future work includes the online implementation of the proposed control framework. The hardware-in-loop (HIL) tests can be conducted to estimate the related energy management controller for online application.